\documentclass[aps,prl,twocolumn,preprintnumbers,superscriptaddress]{revtex4}

\usepackage{color}
\usepackage{multirow}

\usepackage[normalem]{ulem}

\usepackage{amsmath}
\usepackage{enumerate}
\usepackage{amsfonts}
\usepackage{epsfig}

\newcommand{\be}{\begin{equation}}
\newcommand{\ee}{\end{equation}}
\newcommand{\bea}{\begin{eqnarray}}
\newcommand{\eea}{\end{eqnarray}}

\def\le{\left}
\def\ri{\right}

\def\nr{N_{\rm r}}
\def\l{l}

\begin{document}

\title{Tales of Lifshitz Tails}
\title{Instanton Calculus of Lifshitz Tails}

\preprint{MIT-CTP-4364}

\author{Sho Yaida}
\affiliation{Center for Theoretical Physics, Massachusetts Institute of Technology,
Cambridge, MA 02139, USA
}


\begin{abstract}
For noninteracting particles moving in a Gaussian random potential, there exists a disagreement in the literature on the asymptotic expression for the density of states in the tail of the band.
We resolve this discrepancy.
Further we illuminate the physical facet of instantons appearing in replica and supersymmetric derivations with another derivation employing a Lagrange multiplier field.
\end{abstract}
\maketitle

\newpage

Pure crystals have band structures with definite gaps in the energy spectrum.
Impurities attach Lifshitz tails to each band~\cite{Lifshitz}, allowing quasiparticles to be trapped deep in the band gaps~\cite{Anderson}.
Building on work by Halperin and Lax~\cite{HL1}, Zittartz and Langer~\cite{ZL} developed a systematic method for obtaining an asymptotic expression for the density of states deep in the tail of the band
\be
\rho(E)\approx A(E) e^{-B(E)}
\ee
for noninteracting quasiparticles moving in a random potential (see also~\cite{HL2}).
A decade later, Cardy revisited the problem using the replica trick~\cite{Cardy}.
An instanton yielded the same exponential factor $e^{-B(E)}$, but zero-mode counting showed that the way the prefactor $A(E)$ scales with $E$ is different from the one presented in~\cite{ZL}.
To confirm our understanding of these methods, it is important to reconcile the dispute~\cite{Kivelson}.

It turns out that the source of the disagreement lies in a minor algebraic mistake.
Correcting Eq.(5.24) of~\cite{ZL} to
\bea
\Big|{\rm det}\le(\nabla \nabla D\ri)\Big|&=&\Big|{\rm det}\le\{2\int d{\bf x}V({\bf x})\nabla \nabla {\bar V}({\bf x})\ri\}\Big|\\
&\approx&\Big|{\rm det}\le\{2\int d{\bf x}{\bar V}({\bf x})\nabla \nabla {\bar V}({\bf x})\ri\}\Big|\equiv c,\nonumber
\eea
the apparent discrepancy between~\cite{ZL} and~\cite{Cardy} is resolved.
The eventual agreement adds confidence to the use of the replica trick in the nonperturbative regime.

In the remainder of this note, we survey various methods for obtaining the asymptotic forms of Lifshitz tails, illuminating connections among them and tying up loose ends.
We first present a derivation utilizing the method of Lagrange multipliers, building upon classic work by Halperin, Lax, Zittartz, and Langer~\cite{33yrs-scooping}.
It adheres a physical interpretation to the instantons appearing in supersymmetric and replica derivations, which we review next.
Common to all three derivations is the appearance of translational zero modes, but each method treats remaining modes differently.
In particular the supersymmetric method replaces ``-1" bosonic zero modes appearing in the replica limit by the combination of one bosonic zero mode and two fermionic zero modes.

Focusing on the vicinity of a band edge, we consider a system of noninteracting quasiparticles in $d$ spatial dimensions, governed by the Schr{\" o}dinger equation
\be
-\frac{\hbar^2}{2m}\nabla^2 \psi^{V}_n({\bf x})+V({\bf x})\psi^{V}_n({\bf x})=E^{V}_n\psi^{V}_n({\bf x})
\ee
where $V({\bf x})$ is a Gaussian random potential which is assumed to self-average on macroscopic scales. In the thermodynamic limit where the volume of the sample $V_d$ approaches infinity, we can then legitimately estimate the disorder-averaged density of states as
\be
\le[\rho(E)\ri]_{\rm d.a.}=\frac{1}{{\cal N}_{\sharp}}\int\le[{\cal D}V\ri] e^{-\frac{1}{2\gamma}\int d{\bf x}V^2({\bf x})}\rho^{V}(E)
\ee
where
\be
\rho^{V}(E)=\frac{1}{V_{d}}\sum_n \delta(E-E_n^{V}).
\ee
Here, ${\cal N}_{\sharp}\equiv\int\le[{\cal D}V\ri] e^{-\frac{1}{2\gamma}\int d{\bf x}V^2({\bf x})}$ is the normalization constant and $\gamma$ characterizes the strength of the disorder.
We are interested in the asymptotic behavior of $\le[\rho(E)\ri]_{\rm d.a.}$ in the limit of large negative $E$.

To this end, we seek a localizing potential which minimizes the cost $\frac{1}{2\gamma}\int d{\bf x}V^2\le({\bf x}\ri)$ while still holding an eigenfunction with the negative eigenenergy $E$.
Note that for $d<4$ a square-integrable potential always has a unique normalizable ground state~\cite{rigorousQM} and this is the eigenfunction we seek henceforth as, otherwise, we can lower the cost of the potential via simple rescaling.
Through the introduction of a Lagrange multiplier field $\lambda\le({\bf x}\ri)$ and a Lagrange multiplier $\mu_0$, the problem becomes equivalent to the minimization of the cost action
\bea
&&I\le[V\le({\bf x}\ri), \psi\le({\bf x}\ri), \lambda\le({\bf x}\ri), \mu_0\ri]\\
&\equiv&+\frac{1}{2\gamma}\int d{\bf x}V^2\le({\bf x}\ri)\nonumber\\
&&-\frac{1}{\gamma}\int d{\bf x}\lambda\le({\bf x}\ri)\le\{E+\frac{\hbar^2}{2m}\nabla^2-V\le({\bf x}\ri)\ri\}\psi\le({\bf x}\ri)\nonumber\\
&&+\mu_0\le\{\int d{\bf x}\psi^2\le({\bf x}\ri)-1\ri\}.\nonumber
\eea

Extremizing it yields
\be
V\le({\bf x}\ri)=-\lambda\le({\bf x}\ri)\psi\le({\bf x}\ri),
\ee
\be
-\frac{\hbar^2}{2m}\nabla^2 \psi \le({\bf x}\ri)+V\le({\bf x}\ri)\psi\le({\bf x}\ri)=E \psi\le({\bf x}\ri),
\ee
$\int d{\bf x}\psi^2\le({\bf x}\ri)=1$, $\mu_0=0$, and
\be
-\frac{\hbar^2}{2m}\nabla^2 \lambda\le({\bf x}\ri)+V\le({\bf x}\ri)\lambda\le({\bf x}\ri)=E \lambda\le({\bf x}\ri).
\ee
The last equality, combined with the uniqueness of the ground state, dictates that $\lambda\le({\bf x}\ri)=\lambda_0\psi\le({\bf x}\ri)$ with a constant $\lambda_0$. Our problem then morphs into the instanton problem with a single real scalar field, which we know it to have spherically symmetric solutions minimizing the action among all the nontrivial stationary points for $d<4$~\cite{footA}.
Thus we have the cost minimizing solutions
\be\label{pot}
V_0\le({\bf x}\ri)=E f^2\le(\sqrt{\frac{-2mE}{\hbar^2}}|{\bf x}-{\bf x}_0|\ri),
\ee
\be\label{wave}
\psi_0\le({\bf x}\ri)=\sqrt{\frac{-E}{\lambda_0}} f\le(\sqrt{\frac{-2mE}{\hbar^2}}|{\bf x}-{\bf x}_0|\ri)
\ee
where $f({\tilde r})$ satisfies
\be\label{govern}
\frac{d^2 f}{d{\tilde r}^2}+\frac{d-1}{{\tilde r}}\frac{df}{d{\tilde r}}-f+f^3=0
\ee
with $\frac{df}{d{\tilde r}}\Big|_{{\tilde r}=0}=0$, monotonically decreasing toward $0$ as ${\tilde r}\rightarrow\infty$. The normalization condition on $\psi_0$ fixes $\lambda_0=c_{\lambda}(-E)^{1-\frac{d}{2}}\le(\frac{2m}{\hbar^2}\ri)^{-\frac{d}{2}}$ with
\be
c_{\lambda}=\frac{2\pi^{\frac{d}{2}}}{\Gamma\le(\frac{d}{2}\ri)}\int_0^{\infty}d{\tilde r} {\tilde r}^{d-1}f^2.
\ee

Evaluating the cost action for these solutions yields the leading exponential factor $e^{-\frac{a_d}{g}}$ with the dimensionless coupling
\be
g(E)=\gamma(-E)^{\frac{d}{2}-2}\le(\frac{2m}{\hbar^2}\ri)^{\frac{d}{2}}
\ee
and the dimensionless number
\be
a_d=\frac{\pi^{\frac{d}{2}}}{\Gamma\le(\frac{d}{2}\ri)}\int_0^{\infty}d{\tilde r} {\tilde r}^{d-1}f^4.
\ee

To evaluate the subleading prefactor, let us first expand around each saddle as
\be
V-V_0=\sum_{\l=0}^{\infty} \xi_{\l} v_{\l}
\ee
where $v_{\l}$'s are a set of orthonormal functions.
We choose 
\be
v_0=A_{\rm E} f^2\le(\sqrt{\frac{-2mE}{\hbar^2}}|{\bf x}-{\bf x}_0|\ri)
\ee
with $A_{\rm E}=(2a_d)^{-\frac{1}{2}} (-E)^{\frac{d}{4}}\le(\frac{2m}{\hbar^2}\ri)^{\frac{d}{4}}$ so that all the other modes will not change the ground state energy to first order in $\xi_{\l}$. Integration over $\xi_0$, hitting the energy delta function in the density of states, leaves us with the factor
\bea
\le(\sqrt{2\pi\gamma}\frac{\partial E}{\partial \xi_0}\ri)^{-1}&=&\le(\sqrt{2\pi\gamma}\int d{\bf x}\psi_0^2 v_0\ri)^{-1}\nonumber\\
&=&\le(\frac{c_{\lambda}}{2\pi^{\frac{1}{2}}a_d^{\frac{1}{2}}}\ri)g_0^{-\frac{1}{2}}\frac{1}{(-E)}
\eea
where we also took care of the factor coming from ${\cal N}_{\sharp}$.

Next for $i=1,...,d$ we choose
\be
v_i=A_{\rm T} \partial_i V_0
\ee
where $A_{\rm T}=c_{\rm T} (-E)^{\frac{d}{4}-\frac{3}{2}}\le(\frac{2m}{\hbar^2}\ri)^{\frac{d}{4}-\frac{1}{2}}$ with
\be
c_{\rm T}=\le\{\frac{8\pi^{\frac{d}{2}}}{d\times\Gamma\le(\frac{d}{2}\ri)}\int_0^{\infty}d{\tilde r} {\tilde r}^{d-1}f^2\le(\frac{df}{d{\tilde r}}\ri)^2\ri\}^{-\frac{1}{2}}.\ee
They are $d$ translational zero modes and integration over these modes should be traded for integration over the collective coordinates ${\bf x}_0$, sweeping along the saddle submanifold in the field space.
The Jacobian involved in this coordinate transformation is $A_{\rm T}^{-1}$ for each mode as can be seen by comparing changes in the field induced by $\le(\delta{\bf x}_0\ri)_i$ and by $\delta \xi_i$.
After dividing by the volume $V_d$ and again taking ${\cal N}_{\sharp}$ into account, we receive
\be
\le(A_{\rm T}\sqrt{2\pi\gamma}\ri)^{-d}=g_0^{-\frac{d}{2}}\le(\frac{-2mE}{\hbar^2}\ri)^{\frac{d}{2}}\le(2\pi c_{\rm T}^2\ri)^{-\frac{d}{2}}
\ee
from these modes.

Finally integration over all the other modes gives a constant of order 1, worked out in the Appendix.

All in all we find
\be\label{tail}
\le[\rho(E)\ri]_{\rm d.a.}\approx c\times(-E)^{\frac{d}{2}-1}\le(\frac{2m}{\hbar^2}\ri)^{\frac{d}{2}}\le\{g(E)\ri\}^{-\frac{(d+1)}{2}}\times e^{-\frac{a_d}{g(E)}}
\ee
for  $d<4$ in the regime $E\ll-\gamma^{\frac{2}{4-d}}\le(\frac{2m}{\hbar^2}\ri)^{\frac{d}{4-d}}$. An expression for the dimensionless overall constant $c$ is given in the Appendix [cf. Eq.(\ref{nerd-sniped})].

Let us now turn to the supersymmetric derivation~\cite{25yrs-scooping, 22yrs-scooping, SUSY}. First we express the density of state as
\be
\rho^{V}(E)=-\frac{1}{\pi}\lim_{\delta\rightarrow +0}{\rm Im}\le[\frac{1}{V_{d}}\int d{\bf x}G_{\rm R}^{V}({\bf x}, {\bf x}; E+i\delta)\ri].
\ee
The retarded one-particle Green function $G_{\rm R}^{V}({\bf x}, {\bf x}'; E+i\delta)$ can be represented as
\be\label{funrep}
(-i)\frac{\int\le[{\cal D}\phi\ri]\phi({\bf x})\phi({\bf x}') e^{i S_V[\phi]}}{\int\le[{\cal D}\phi\ri] e^{i S_V[\phi]}}
\ee
with
\be
S_V[\phi]=\frac{1}{2}\int d{\bf x}\phi\le\{E+i\delta+\frac{\hbar^2}{2m}\nabla^2-V\ri\}\phi.
\ee
The supersymmetric method proceeds by rewriting the expression (\ref{funrep}) as
\be\label{SUSYrep}
\le(\frac{-i}{2}\ri)\int\le[{\cal D}{\vec \phi}{\cal D}\chi_1{\cal D}\chi_2\ri]{\vec \phi}({\bf x})\cdot {\vec \phi}({\bf x}') e^{\sum_{a=1}^2 i S_V[\phi_a]+i S_V[\chi_1, \chi_2]}
\ee
with
\be
S_V[\chi_1, \chi_2]=\frac{1}{2}\int d{\bf x}\chi_2\le\{E+i\delta+\frac{\hbar^2}{2m}\nabla^2-V\ri\}\chi_1
\ee
where we doubled the bosonic field $\phi$ to ${\vec \phi}=(\phi_1, \phi_2)$ and introduced fermionic fields $\chi_1$ and $\chi_2$.
Now that there is no denominator containing the random potential, we can perform functional integration over $V$ and obtain
\bea\label{anal}
\le[\rho(E)\ri]_{\rm d.a.}&=&\frac{1}{2\pi V_{d}}{\rm Im}\int\le[{\cal D}{\vec {\tilde \phi}}{\cal D}{\tilde \chi}_2{\cal D}{\tilde \chi}_1\ri]\int d{\bf x}{\vec {\tilde \phi}}({\bf x})\cdot {\vec {\tilde \phi}}({\bf x})\nonumber\\
&&\times e^{-S_{\rm b}[{\vec {\tilde \phi}}]-S_{2{\rm f}}[{\tilde \chi}_1, {\tilde \chi}_2, {\vec {\tilde \phi}}]-S_{4{\rm f}}[{\tilde \chi}_1, {\tilde \chi}_2]}
\eea
with
\be
S_{\rm b}[{\vec {\tilde \phi}}]=\frac{1}{2}\int d{\bf x}\le[{\vec {\tilde \phi}}\cdot\le(-\frac{\hbar^2}{2m}\nabla^2-E-\frac{\gamma}{4}{\vec {\tilde \phi}}^2\ri){\vec {\tilde \phi}}\ri],
\ee
\be
S_{2{\rm f}}[{\tilde \chi}_1, {\tilde \chi}_2, {\vec {\tilde \phi}}]=\frac{1}{2}\int d{\bf x}{\tilde \chi}_2\le(-\frac{\hbar^2}{2m}\nabla^2-E-\frac{\gamma}{2}{\vec {\tilde \phi}}^2\ri){\tilde \chi}_1,
\ee
and
\be
S_{4{\rm f}}[{\tilde \chi}_1, {\tilde \chi}_2]=-\frac{\gamma}{8}\int d{\bf x}\le({\tilde \chi}_2{\tilde \chi}_1\ri)^2.
\ee
We have defined ${\vec {\tilde \phi}}\equiv e^{+\frac{i\pi}{4}}{\vec \phi}$ and ${\tilde \chi}_a\equiv e^{+\frac{i\pi}{4}}\chi_a$ for $E<0$ and the expression (\ref{anal}) should be viewed with appropriate analytic continuation in mind~\cite{Langer, Coleman}.

To evaluate $\le[\rho(E)\ri]_{\rm d.a.}$ for large negative $E$, we use the method of steepest descent, extremizing  $S_{\rm b}$.
The trivial saddle ${\vec {\tilde \phi}}=0$ gives no contribution to $\le[\rho(E)\ri]_{\rm d.a.}$ due to the absence of negative modes. 
Among nontrivial saddles, we assume~\cite{footB} that the saddles
\be\label{instanton}
{\vec {\tilde \phi}}_{\rm cl}\le({\bf x}\ri)={\vec e}\sqrt{\frac{-2E}{\gamma}} f\le(\sqrt{\frac{-2mE}{\hbar^2}}|{\bf x}-{\bf x}_0|\ri)
\ee
minimize the action, where ${\vec e}$ is a constant unit vector and $f({\tilde r})$ was defined around Eq.(\ref{govern}).
Evaluating the action for these solutions gives the same leading exponential factor $e^{-\frac{a_d}{g}}$ as before.

In regards to the subleading prefactor, one contribution comes from
\be
\int d{\bf x}{\vec {\tilde \phi}}_{\rm cl}({\bf x})\cdot {\vec {\tilde \phi}}_{\rm cl}({\bf x})\sim g_0^{-1}(-E)^{-1}
\ee
in front.
(We will not keep track of the overall dimensionless constant in this derivation.)
To evaluate the remaining contributions, we expand around each saddle as
\be
{\vec {\tilde \phi}}-{\vec {\tilde \phi}}_{\rm cl}={\vec e}\sum_{\l=0}^{\infty} \xi^{{\rm B}\parallel}_{\l} \varphi^{{\rm B}\parallel}_{\l}+{\vec e}_{\perp}\sum_{\l=0}^{\infty} \xi^{{\rm B}\perp}_{\l} \varphi^{{\rm B}\perp}_{\l}
\ee
and
\be
{\tilde \chi}_a=\sum_{\l=0}^{\infty}\le(\xi^{{\rm F}}_\l\ri)_a \varphi^{{\rm F}}_{\l}.
\ee
Here, ${\vec e}_{\perp}$ is a unit vector perpendicular to ${\vec e}$, $\varphi^{{\rm B}\parallel}_{\l}$'s are a set of orthonormal functions satisfying
\be
\le(-\frac{\hbar^2}{2m}\nabla^2-E-\frac{3\gamma}{2}{\vec {\tilde \phi}}^2_{\rm cl}\ri)\varphi^{{\rm B}\parallel}_{\l}=(-E)c_{\l}^{\parallel}\varphi^{{\rm B}\parallel}_{\l},
\ee
and $\varphi^{{\rm B}\perp}_{\l}=\varphi^{{\rm F}}_{\l}\equiv\varphi^{\perp}_{\l}$'s are another set of orthonormal functions satisfying
\be
\le(-\frac{\hbar^2}{2m}\nabla^2-E-\frac{\gamma}{2}{\vec {\tilde \phi}}^2_{\rm cl}\ri)\varphi^{\perp}_{\l}=(-E)c_{\l}^{\perp}\varphi^{\perp}_{\l},
\ee
with dimensionless numbers $c_{\l}^{\parallel}$'s and $c_{\l}^{\perp}$'s. We deal first with $\xi^{{\rm B}\parallel}_{\l}$ fluctuations and then with the rest.

Analyzing $\varphi^{{\rm B}\parallel}_{\l}$ modes, we find that the lowest mode has a negative eigenvalue $c_0^{\parallel}<0$, giving rise to a factor of $i(-E)^{-\frac{1}{2}}$ and allowing the saddles to contribute to the density of states.
Next come $d$ translational zero modes.
Trading integration over these modes for integration over ${\bf x}_0$ and dividing by the volume, we receive the Jacobian
\be
\le\{\frac{1}{d}\int d{\bf x}\le(\nabla {\vec {\tilde \phi}}_{\rm cl}\ri)^2\ri\}^{\frac{d}{2}}\sim (-E)^{-\frac{d}{2}}\le\{g_0^{-\frac{d}{2}}\le(\frac{-2mE}{\hbar^2}\ri)^{\frac{d}{2}}\ri\}.
\ee
Finally all the other modes have positive eigenvalues, each of which gives a factor of $(-E)^{-\frac{1}{2}}$.

Analyzing the other set of fluctuations, except the lowest modes, all the modes have positive eigenvalues, each of which gives a factor of $(-E)^{-\frac{1}{2}+\frac{1}{2}+\frac{1}{2}}=(-E)^{+\frac{1}{2}}$. The lowest modes are the zero modes arising from $O(2)$-rotational symmetry, proportional to $\Big|{\vec {\tilde \phi}}_{\rm cl}\le({\bf x}\ri)\Big|$.
The bosonic zero mode, upon trading integration over $\xi^{{\rm B}\perp}_0$ for integration over ${\vec e}$, yields the Jacobian
\be
\le\{\int d{\bf x} {\vec {\tilde \phi}}_{\rm cl}^2\ri\}^{\frac{1}{2}}\sim (-E)^{-\frac{1}{2}}\le(g_0^{-\frac{1}{2}}\ri).
\ee
We also need to saturate fermionic zero modes by expanding the action to the quartic order in fluctuations:
if we kept only quadratic terms in the expansion of the action, integration over $\le(\xi^{{\rm F}}_0\ri)_a$'s would give zero.
Thus we must bring down either a factor of $\gamma\int d{\bf x}{\tilde \chi}_2\le({\vec {\tilde \phi}}^2-{\vec {\tilde \phi}}_{\rm cl}^2\ri){\tilde \chi}_1$ or $\gamma\int d{\bf x}\le({\tilde \chi}_2{\tilde \chi}_1\ri)^2$. After appropriate Gaussian integrations, we obtain a factor of
\be
(-E)^{+\frac{2}{2}}\le(g_0^{+\frac{2}{2}}\ri).
\ee

Putting them all together, we recover the same result (\ref{tail}) as before.

Finally let us turn to the replica derivation~\cite{Cardy}.
The replica trick proceeds by rewriting the expression (\ref{funrep}) as
\be
\le(\frac{-i}{\nr}\ri)\int\le[{\cal D}{\vec \phi}\ri]{\vec \phi}({\bf x})\cdot {\vec \phi}({\bf x}') e^{i \sum_{a=1}^{\nr}S_V[\phi_a]}
\ee
where we introduced $(\nr-1)$ replicas, promoting $\phi$ to ${\vec \phi}=(\phi_1, \phi_2,..., \phi_{\nr})$, and took
the dicey limit in which $\nr\rightarrow 0$ to eliminate the denominator. After integrating over $V$ and making analytic continuation, we find instantons of the same form (\ref{instanton}), but zero-mode analysis is slightly different from the one in the supersymmetric derivation. Besides $d$ translational zero modes, there are $\lim_{\nr\rightarrow0}(\nr-1)=-1$ bosonic zero modes coming from $O(\nr)$-rotational symmetry. 
The latter is replaced by the combination of one bosonic zero mode and two fermionic zero modes in the supersymmetric derivation.

We note that the instantons (\ref{instanton}) appearing in replica and supersymmetric derivations and the localized wavefunctions (\ref{wave}) have exactly the same shape.
Thus we interpret the instantons as most likely forms of localized wavefunctions or square roots of localizing potentials [cf. Eq.(\ref{pot})], dilutely distributed for large negative $E$.
It may be more appropriate to call all these solutions ``localons."

Through the use of the Lagrange multiplier field, we simplified the method developed by Halperin, Lax, Zittartz, and Langer.
Instantons appearing in replica and supersymmetric derivations are now interpreted as localized wavefunctions or as square roots of localizing potentials.
All three derivations are shown to give the same asymptotic expression for the Lifshitz tail in the case of the Gaussian random potential, including the subleading prefactor.

The author thanks Steven~A.~Kivelson for pointing out the discrepancy between~\cite{ZL} and~\cite{Cardy} and for emphasizing the conceptual importance of the would-be conflict.
He thanks Allan~W.~Adams, John~L.~Cardy, John~A.~McGreevy, and Stephen~H.~Shenker for comments, discussions, and encouragement.
He is supported by a JSPS Postdoctoral Fellowship for Research Abroad.

\appendix

\section{APPENDIX}
For $\l \geq d+1$, we have the ground state energy shift
\be
\sum_{\l=d+1}^{\infty}\sum_{\l'=d+1}^{\infty} \xi_{\l}\xi_{\l'}\sum_{n}\frac{\langle 0|v_{\l}|n\rangle\langle n|v_{\l'}|0\rangle}{E_0^{V_0}-E_n^{V_0}}
\ee
to second order in $\xi_{\l}$'s. We compensate it by setting
\be
\xi_0=-\frac{1}{{\langle 0|v_0|0\rangle}}\sum_{\l=d+1}^{\infty}\sum_{\l'=d+1}^{\infty} \xi_{\l}\xi_{\l'}\sum_{n}\frac{\langle 0|v_{\l}|n\rangle\langle n|v_{\l'}|0\rangle}{E_0^{V_0}-E_n^{V_0}}
\ee
so as to keep the ground state energy intact to this order. The resulting disorder cost is
\be
\frac{1}{2\gamma}\sum_{\l=d+1}^{\infty}\sum_{\l'=d+1}^{\infty} \xi_{\l}\xi_{\l'}\le(\delta_{l, l'}-2\lambda_0\sum_{n}\frac{\langle 0|v_{\l}|n\rangle\langle n|v_{\l'}|0\rangle}{E_n^{V_0}-E}\ri).
\ee
Imitating~\cite{ZL}, we proceed by choosing $v_{\l}=f u_{\l}$ with
\be
\le[-\frac{\hbar^2}{2m}\nabla^2-E+(1+c_{\l})V_0\ri]u_{\l}=0.
\ee
(Corresponding to $v_0$ and $v_i$, we have $u_0\propto f$ and $u_i\propto \partial_i f$ with $c_{0}=0$ and $c_i=2$, respectively.) With this trick we evaluate the cost to be
\be
\frac{1}{2\gamma}\sum_{\l=d+1}^{\infty} \xi_{\l}^2\le(1-\frac{2}{c_{\l}}\ri).
\ee
Performing Gaussian integrals over $\xi_l$'s for $\l \geq d+1$, taking ${\cal N}_{\sharp}$ into account, and combining with the contributions from the other modes, we obtain
\be\label{nerd-sniped}
c=\le(\frac{c_{\lambda}}{2\pi^{\frac{1}{2}}a_d^{\frac{1}{2}}}\ri) \le(2\pi c_{\rm T}^2\ri)^{-\frac{d}{2}}\prod_{\l=d+1}^{\infty}\le(1-\frac{2}{c_{\l}}\ri)^{-\frac{1}{2}}.
\ee
Another expression for $c$ is given in~\cite{Houghton}.

For $d=1$, we get $a_d=\frac{8}{3}$, $c_{\lambda}=4$, $c_{\rm T}=\frac{\sqrt{15}}{8}$, and $c_{\l}=\frac{l(l+3)}{2}$, the last of which can be obtained through the use of Gegenbauer polynomials of order $\frac{3}{2}$~\cite{ZL}.
Thus $c=\frac{4}{\pi}$, conforming with the exact result obtained by Halperin~\cite{brutal}.


\begin{thebibliography}{11}
\bibitem{33yrs-scooping}
We perform the restricted functional integral in a way different from the one presented in previous work~\cite{Houghton}.
There they introduce a Lagrange multiplier while here we introduce a Lagrange multiplier field.
\bibitem{footA}
  The case for $d=3$ is worked out in~\cite{CGM} and the case for $d=2$ is worked out in~\cite{BGK, BL}.
  For $d=1$, $f({\tilde r})=\frac{\sqrt{2}}{{\rm cosh}({\tilde r})}$ gives a desired solution [cf. Eq.(\ref{govern})].
\bibitem{footB}
  For $d=3$, it has been argued that the saddles of the form (\ref{instanton}) have the least action among all the nontrivial stationary points~\cite{LT}, extending the theorem in~\cite{CGM}.
  For $d\leq2$, the author is unaware of analogous extension to multiple fields needed in replica and supersymmetric derivations, but the resultant agreement with the other derivaton strongly indicates that they indeed minimize the action among all the nontrivial saddles.
\end{thebibliography}
\end{document}